\documentclass[aps,twocolumn,superscriptaddress, prb, nofootinbib]{revtex4-1}
\usepackage{amsmath}
\usepackage{graphicx}

\begin{document}	
\title{Elucidating the Structure of the Magnesium Aluminum Chloride Complex electrolyte for
Magnesium-ion batteries}

\author{Pieremanuele Canepa} \email{pcanepa@mit.edu; pcanepa@lbl.gov}
\affiliation{
Department of Materials Science and Engineering, Massachusetts Institute
of Technology, Cambridge, MA 02139, USA.
}
\affiliation{
Materials Science Division, Lawrence Berkeley National Laboratory,
Berkeley, CA 94720, USA.  
}

\author{Saivenkataraman Jayaraman}
\affiliation{
Department of Materials Science and Engineering, Massachusetts Institute
of Technology, Cambridge, MA 02139, USA.
}
\altaffiliation{Current Affiliation:The Chemours Company, 1007 N Market Street,
Wilmington, DE 19899, USA}

\author{Lei Cheng}
\affiliation{
Materials Science Division, Argonne National Laboratory, Argonne, IL
60439, USA. 
}

\author{Nav Nidhi Rajput}
\affiliation{
Environmental Energy Technologies Division, Lawrence Berkeley National
Laboratory, Berkeley, CA 94720, USA.  }

\author{William D.\ Richards}
\affiliation{
Department of Materials Science and Engineering, Massachusetts Institute
of Technology, Cambridge, MA 02139, USA.
}

\author{Gopalakrishnan Sai Gautam}
\affiliation{
Department of Materials Science and Engineering, Massachusetts Institute
of Technology, Cambridge, MA 02139, USA.
}
\affiliation{
Materials Science Division, Lawrence Berkeley National Laboratory,
Berkeley, CA 94720, USA.  
}

\author{Larry A.\ Curtiss}
\affiliation{
Materials Science Division, Argonne National Laboratory, Argonne, IL
60439, USA. 
}

\author{Kristin A.\ Persson}
\affiliation{
Environmental Energy Technologies Division, Lawrence Berkeley National
Laboratory, Berkeley, CA 94720, USA.  }

\author{Gerbrand Ceder}\email{gceder@lbl.gov; gceder@berkeley.edu}
\affiliation{
Department of Materials Science and Engineering, University of California Berkeley, Berkeley, CA 94720, USA.
}
\affiliation{
Materials Science Division, Lawrence Berkeley National Laboratory,
Berkeley, CA 94720, USA.  
}
\affiliation{
Department of Materials Science and Engineering, Massachusetts Institute
of Technology, Cambridge, MA 02139, USA.
} 

\begin{abstract}
\noindent {\bf Broader context:}
Electrical energy storage is a key technology for a clean energy economy,
but currently requires significant improvement in energy density beyond
the capabilities of traditional Li-ion batteries. Mg-ion batteries offer
an exciting alternative in terms of the amount of energy that can be
delivered, safety, manufacturing and disposal costs, with limited
environmental impact.  The electrochemical functions of the Mg-ion
battery ultimately depend on the choice of the electrolyte, which is
limited by the peculiar chemistry of Mg. To date, very few electrolytes
can reversibly plate and strip Mg. The Magnesium Aluminum Chloride
Complex (MACC) electrolyte can reversibly plate and strip Mg with
significantly higher voltage (3.1 V) as compared to other electrolytes,
but there is a pressing need to address critical questions about the
structural evolution of this electrolyte during electrochemical cycling.
\break
{\bf Abstract:} Non-aqueous Mg-ion batteries offer a promising way to
overcome safety, costs, and energy density limitations of
state-of-the-art Li-ion battery technology. We present a rigorous
analysis of the Magnesium Aluminum Chloro Complex (MACC) in
tetrahydrofuran (THF), one of the few electrolytes that can reversibly
plate and strip Mg. We use \emph{ab initio} calculations and
classical molecular dynamics simulations to interrogate the MACC
electrolyte composition with the goal of addressing two urgent questions
that have puzzled battery researchers: \emph{i}) the functional species
of the electrolyte, and \emph{ii}) the complex equilibria regulating the
MACC speciation after prolonged electrochemical cycling, a process
termed as conditioning, and after prolonged inactivity, a process called
aging. A general computational strategy to untangle the complex
structure of electrolytes, ionic liquids and other liquid media is
presented. The analysis of formation energies and grand-potential phase
diagrams of Mg-Al-Cl-THF suggests that the MACC electrolyte bears a
simple chemical structure with few simple constituents, namely the
electro-active species MgCl$^+$ and AlCl$_4^-$ in equilibrium with
MgCl$_2$ and AlCl$_3$. Knowledge of the stable species of the MACC
electrolyte allows us to determine the most important equilibria
occurring during electrochemical cycling. We observe that Al deposition
is always preferred to Mg deposition, explaining why freshly synthesized
MACC cannot operate and needs to undergo preparatory conditioning.
Similarly, we suggest that aluminum displacement and depletion from the
solution upon electrolyte resting (along with continuous MgCl$_2$
regeneration) represents one of the causes of electrolyte aging.
Finally, we compute the NMR shifts from shielding tensors of selected
molecules and ions providing fingerprints to guide future experimental
investigations.
\end{abstract}

\maketitle

\section{Introduction}
\label{sec:intro}

The success of clean energy sources is predicated on improvements in
energy storage technologies.  State-of-the-art Li-ion batteries,
although instrumental in considerable advances in portable electronics,
cannot cope with the minimum
storage\cite{VanNoorden2014,Yoo2013,Muldoon2014,Shterenberg2014} and
safety\cite{Cohen2000} requirements dictated by grid and transport
applications. 

A viable strategy for  post-Li-ion technology is to replace  Li with
safer and  earth-abundant  Mg. Magnesium  has the advantage of doubling
the total charge per ion, which results in larger theoretical volumetric
capacity compared to typical Li-ion
batteries.\cite{Aurbach2002,Muldoon2012,Yoo2013,VanNoorden2014,Muldoon2014,Shterenberg2014}
Most importantly, in Mg-ion batteries the intercalation architecture of
the graphitic-anode for Li-ions is replaced by a high-energy density
metal anode ($\sim$~700~Ah~l$^{-1}$ and $\sim$~3830 Ah~l$^{-1}$,
respectively).\cite{Aurbach2002,Muldoon2012,VanNoorden2014,Muldoon2014,Shterenberg2014,Liu2015}

Notwithstanding the tantalizing advantages of Mg-ion technology, its
distinct electrochemistry imposes serious limitations on the kind of
electrolyte that can reversibly plate and strip Mg, and at the same time
sustain high-voltage cathode materials. For example, Mg-ion electrolytes
that are analogous to their Li-ion counterparts (e.g.\ PF$_6^-$ Li$^+$)
and solvents (e.g.\ propylene carbonate/dimethyl carbonate) irreversibly
decompose at the Mg anode, producing passivating layers that are
impermeable to Mg-ions, and inhibit further electrochemical
activity.\cite{Brow1985,Aurbach2011}

To circumvent this issue, efforts by Gregory \emph{et
al.}\cite{Gregory1990,Muldoon2014} demonstrated \emph{quasi}-reversible
Mg-plating from  Grignard's reagents. Greater coulombic efficiencies and
anodic stabilities were achieved by Aurbach and collaborators after many
years of meticulous tuning of the organic magnesium aluminum chloride
salts (organo-magnesium-chloride complexes) dissolved in ethereal
solutions, namely the dichloro complex  (DCC) and the ``all phenyl
complex''
(APC).\cite{Aurbach2001,Aurbach2002a,Gizbar2004,Mizrahi2008,Aurbach2011,Pour2011,Yoo2013,Doe2013,Doe2014,Barile2014a,Barile2014}
Similarly, Shao \emph{et al.}\cite{Shao2013} achieved Mg deposition by
combining Mg(BH$_4$)$_2$ and LiBH$_4$ in diglyme. The air-sensitivity
and low anodic stability of previous Mg-ion electrolytes led Kim
\emph{et al.}\cite{Kim2011}  to propose a non-nucleophilic salt
comprising AlCl$_3$ and hexamethyldisilazide magnesium chloride
(HMDSMgCl). Recently, Mohtadi, Arthur and co-workers at Toyota developed a series of halogen--free electrolyte based on Mg borohydride, boron-clusters and carboranes, which are not corrosive and with relatively high anodic-stability $\sim$~3.8--4.0~V.\cite{Tutusaus2015,Carter2014,Mohtadi2012} Subsequently, Doe~\emph{et al.}\cite{Doe2013,Doe2014}
developed an inexpensive electrolyte termed Magnesium Aluminum Chloride
Complex (MACC) which is formed by mixing two common inorganic salts,
namely AlCl$_3$ with MgCl$_2$ in ethereal solutions. MACC possesses a
relatively large anodic stability ($\sim$~3.1~V) and good reversible Mg
deposition/stripping.  The MACC electrolyte is the focus of the current
paper. 

The good performance of an electrolyte is dictated by few but important
parameters such as high coulombic efficiency, high anodic and cathodic
stabilities, and high diffusivity of the ion carriers, which depend
ultimately on the structural composition of the electrolyte at rest and
during electrochemical cycling. In the present study we interrogate the
MACC electrolyte composition with the goal of elucidating: \emph{i}) the
functional species of the electrolyte, and \emph{ii}) the complex
equilibria regulating the MACC speciation after prolonged
electrochemical cycling, a process termed as
conditioning\cite{Shterenberg2014}, and after inactivity, termed as
aging\cite{Barile2014}. Aiming to describe important macroscopic effects
observed in electrochemical experiments from the ground-up, and provide
an atomistic picture of the processes regulating the speciation in the
MACC electrolyte at different electrochemical conditions, we explore the
complex chemical space of Mg-Cl-Al-THF (with THF as
tetrahydrofuran) combining Density Functional Theory (DFT) with
Classical Molecular Dynamics (CMD) simulations. The computational
strategy is general and suitable to other electrolytes, ionic liquids
and a variety of liquid media. 

On the basis of previous experimental XRD, Raman and NMR
observations\cite{Pour2011} on the APC electrolyte, similar to MACC but
with organic moieties on the Al$^{3+}$ ion, it is speculated that ${\rm
Mg}_m{\rm Al}_n{\rm Cl}_{[(2m)+(3n)]}$ comprises the magnesium chloride
monomer, $[\mu-$MgCl$\cdot$5THF$]^+$,  the dimer
$[\mu-$Mg$_2$Cl$_3\cdot$6THF$]^+$ and AlCl$_4^-$ as counterion. While
monomer and dimer ions are yet to be observed during electrochemical
cycling in both MACC and APC, they are thought to be the active
Mg$^{2+}$ carriers during electrochemical cycling.  Barile \emph{et
al.}\cite{Barile2014} also speculated that higher order magnesium-chloro
structures such as trimer and multimeric units may exist, and they are
included in this study.

With the aid of density functional theory calculations and molecular
dynamics we are able to show that the MACC electrolyte bears a simple
chemical structure with very few species present. By knowing the stable
species of the electrolyte, we elucidate the equilibria taking place in
the electrolyte, showing that the Mg$^{2+}$ carriers, MgCl$^+$ are
continuously exchanged with MgCl$_2$ (and AlCl$_4^-$ with AlCl$_3$), and
changes of these equilibria alter the observed electrochemical
performance of the electrolyte.  Finally, our results suggest some
explanation for the phenomena of electrolyte aging and conditioning. 

These findings are instrumental for progressing the development of the
next generation of Mg-ion batteries. Particularly, we demonstrate the
working of the MACC electrolyte, and also provide clear directions for
the improvement of the electrolyte performance.

\section{Methodology}
\label{sec:method}

\subsection{\emph{Ab initio} molecular solvation and periodic bulk calculations}
\label{subsec:QM}
\begin{figure}[ht]
\centering
\includegraphics[width=1.0\columnwidth]{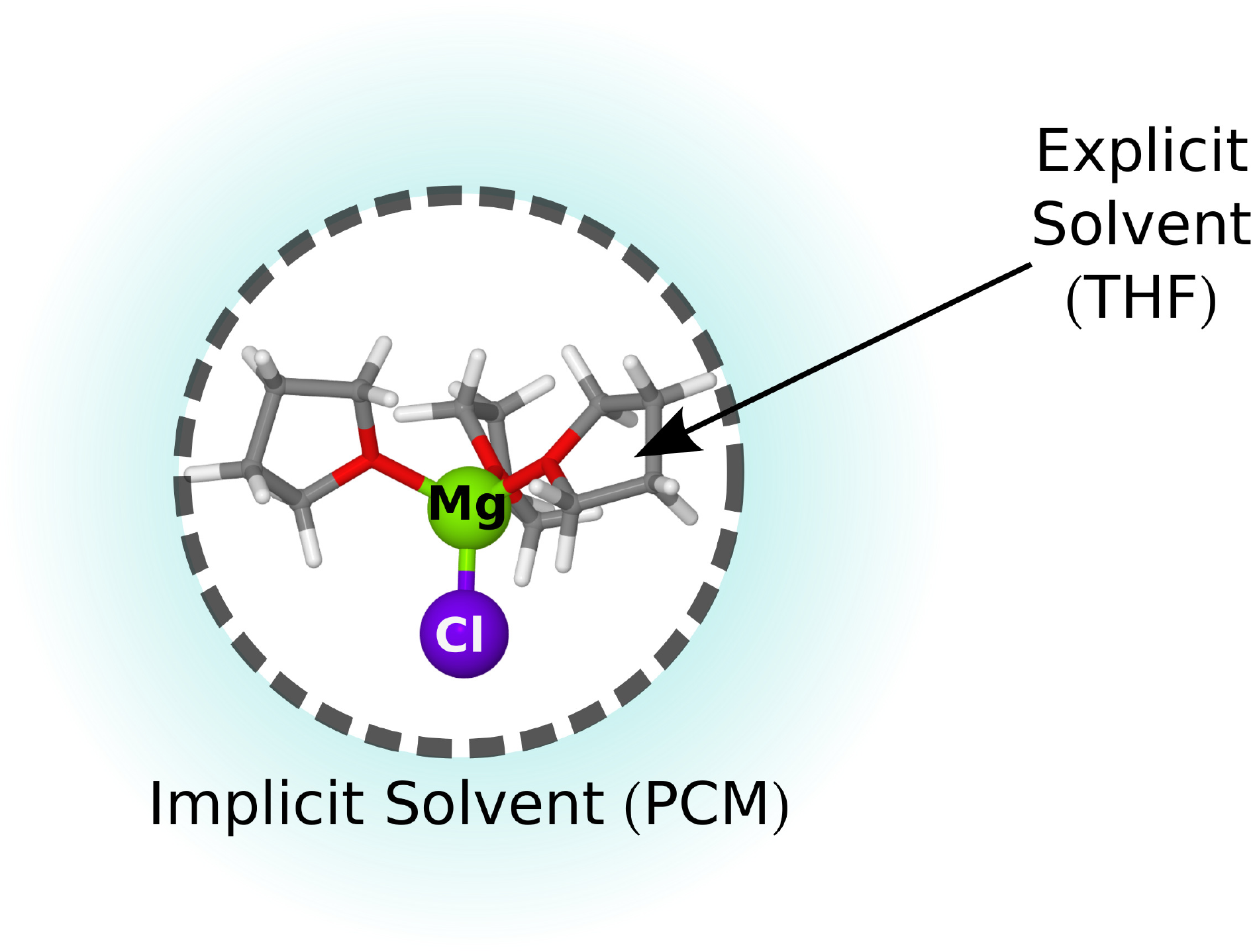}
\caption{\label{fig:solvation} Diagram showing the approximation used to
capture the solvation structure  of magnesium-aluminum-chloride
complexes (see Equations~\ref{eq:foamtionenergy}).  Inner circle MACC
cluster with a 1$^{st}$ shell of explicit solvent of THFs, outer circle
(cyan halo) for longer range solvation shells.}
\end{figure}

Due to the importance of strong interactions between solvent molecules
and the species in the MACC electrolyte, we model, using Density
Functional Theory (DFT), each magnesium-aluminum-chloride complex with
an explicit solvent of THF molecules in the first solvation shell and an
implicit model in the outer shells to describe long-range
solvent-solvent interactions. The methodology is shown in
Fig.~\ref{fig:solvation}, where 3 THFs are included in the first
solvation shell of MgCl$^+$, while the domain indicated by the cyan halo
depicts outer shells, which are modeled as a dielectric medium by the
Polarizable Continuum Model (PCM).\cite{Tomasi2005}

According to Figure~\ref{fig:solvation} $G_{\rm PCM}$, the Gibbs free
energy of the fully solvated Mg-Cl-Al-THF clusters are set by
Equation~\ref{eq:foamtionenergy}.
\begin{equation}
\label{eq:foamtionenergy}
G_{\rm PCM} = E_{\rm PCM} + {\rm ZPE}_{\rm expl} + q_{\rm expl} -
TS_{\rm expl}
\end{equation}
where $G_{\rm PCM}$ and $E_{\rm PCM}$ represent the Gibbs free energy
and total energy of the fully solvated complex (explicit solvent and
implicit solvent, see Figure~\ref{fig:solvation}), while ZPE$_{\rm
expl}$, $q_{\rm expl}$, and $TS_{\rm expl}$  are the zero point energy
correction, the thermal contribution, and the entropic term
respectively, approximated by the MACC clusters comprising only the
explicit solvent.  To obtain the explicit part of $E_{\rm PCM}$ (see
Fig.~\ref{fig:solvation}), we first relax the geometries of the
magnesium-aluminum-chloro clusters (comprising an explicit 1$^{st}$
solvation shell of THFs) within the DFT approximation with B3LYP and a
6-31+G(d) basis-set implemented in Gaussian09.\cite{g09} Previous
battery studies have demonstrated that B3LYP can accurately reproduce
experimental results.\cite{Rajput2015,Ong2011,Zhang2014,Okoshi2013}
More details on the methodology are provided in the Supplementary
information. Finally, $E_{\rm PCM}$ of Eq.~\ref{eq:foamtionenergy} is
obtained from a single point energy calculation on the relaxed
structures using the PCM model.\cite{Tomasi2005} Frequency analysis is
performed to ensure that the relaxed structures are real minima as well
as to compute the free energy corrections (i.e.\ ZPE$_{\rm expl}$,
$q_{\rm expl}$, and $TS_{\rm expl}$).

To evaluate some important reactions involving solid phases, we employ
the B3LYP functional available in VASP.\cite{Kresse93,Kresse96} The
total energy is sampled on a well-converged
4$\times$4$\times$4~\emph{k}-point grid (and a
16$\times$16$\times$16~\emph{k}-point grid for Al and Mg metals)
together with projector augmented-wave theory\cite{Kresse99} and a 520
eV plane-wave cutoff.  Forces on atoms are converged to less than
1$\times$10$^{-2}$ eV/{\AA}.   In order to compare the  liquid species
participating to chemical reactions containing solid phases (see
Table~\ref{tb:equil_reactions}) we  simulated  the relaxed structures
obtained from molecular PCM calculations with periodic boundary
conditions employing a box of size 20$\times$20$\times$20~\AA$^3$ and
the VASP setup indicated above. Thus, the chemical potential of
molecules and ions coordinated by THF (see
Table~\ref{tb:equil_reactions}) are referenced to the liquid THF via the
experimental enthalpy of vaporization $\Delta$H~$\sim$~0.331
eV.\cite{Scott1970} 

For relevant clusters, $^{35}$Cl and $^{25}$Mg NMR shielding tensors
(only the isotropic shielding part is discussed) are provided as useful
fingerprints to guide experiments. NMR parameters are obtained with
Gauge-Independent Atomic Orbital theory\cite{Wolinski1990} on the
relaxed structures at the 6-31+G(d) level of accuracy, but increasing
the basis-set quality to 6-311+G(d,p). Basis-set convergence on the NMR
isotropic shielding for these molecules are discussed in the
Supplementary Information.

\subsection{Debye-H{\"{u}}ckel correction}
To account for the electrostatic interactions  of ions in the
electrolytic solution, we apply a potential energy correction to the
reactions energies (see Table~{\ref{tb:equil_reactions}}) based on
Debye-H\"{u}ckel theory.{\cite{Debye1923,Robinson1968}} Therefore the
$\Delta$E corrected by the Debye-H{\"{u}}ckel model $\Delta$E$_{D-H}$
becomes:
\begin{equation}
\Delta E_{D-H} = \Delta E + \sum^m_{i = 0} u_i
\end{equation}
where $\Delta$E  obtained from DFT calculations at infinite
dilution,  $m$ is the total number of ion $i$, and $u_i$  the
electrostatic potential energy given by
Equation~{\ref{eq:debye_huckel}}, 

\begin{equation}
\label{eq:debye_huckel}
u_i = -\frac{z_i^2 e^2 \kappa}{8 \pi \epsilon_r \epsilon_0} \frac{1}{1 +
\kappa a_0}
\end{equation}

\begin{equation}
\label{eq:screening_length}
\kappa^2 = \sum_{i} \frac{z_i^2 e^2 c_i^0}{\epsilon_r \epsilon_0 k_B T}
\end{equation}
where $z_i$ is the charge number and $c_i^0$ the number
concentration of ion $i$, $\epsilon_r$ the relative dielectric constant
($\sim$7.5 for THF), $\epsilon_0$ the vacuum permittivity, $k_B$ the
Boltzmann constant, $T$ the temperature, $e$ the electron charge, $a_0$
the minimum separation of ions, and $\kappa^{-1}$ the Debye screening
length.  We set $a_0$ to be $\sim$~7.1~{\AA} that is the minimum
separation of the van der Waals spheres of MgCl$^+$(3THF) and
AlCl$_4^-$.  Since the $\Delta$E$_{D-H }$ of
Equation~{\ref{eq:debye_huckel}} depends on the ionic activity $c_i^0$,
which in turn depends on the magnitude of the Debye-H\"uckel correction,
the $\Delta$E$_{D-H }$  has to be  evaluated numerically through an
iterative  self-consistent procedure. Self-consistency of $u_i$  is
achieved when the concentration  (of the charged species, i.e. MgCl$^+$)
equals the input concentration. 
In general, the Debye-H\"{u}ckel theory is not appropriate for the description of concentrated 
solutions; for this reason we use the extended  Debye-H\"{u}ckel approximation (which holds for concentrations  $<$ 10$^{-1}$~M), see Eq.{~\ref{eq:debye_huckel}}  which is compatible with the concentrations of the charged species in solution ($\sim$ 92 mM for MgCl$^+$ and AlCl$_4^-$).


\subsection{Classical molecular dynamics simulations of bulk electrolytes}

All classical molecular dynamics (CMD) simulations to study the dynamic
structure of the MACC electrolyte are computed using
LAMMPS\cite{LAMMPSref} and treat the effect of the THF solvent
explicitly. The THF-THF, and THF-ion interactions are modeled using the
Generalized Amber Force Field~\cite{Wang2004,Wang2006} (GAFF), whereas
Mg and Cl partial charges presented in Table~\ref{tb:charges} are
computed with the RESP procedure by fitting the electrostatic potential
surface of the optimized geometries using
Antechamber.\cite{Bayly1993,Wang2004,Wang2006}
\begin{table}[th]
\caption{\label{tb:charges} Computed RESP charges and van der Waals
parameters ($\epsilon$~in~kcal~mol$^{-1}$ and $\sigma$~in~\AA) for Mg
and Cl used in the classical CMD simulations of MACC electrolyte.}
\begin{tabular*}{\columnwidth}{@{\extracolsep{\fill}}lcc@{}}
\hline\hline
Species                & Mg          &  Cl          \\
\hline 
MgCl$_2$               & 0.9380      & --0.4690      \\ 
MgCl$^+$, Monomer      & 1.4021      & --0.4021$^*$  \\
Mg$_2$Cl$_3^+$, Dimer  & 1.2621      & --0.5081$^*$  \\
Mg$_3$Cl$_5^+$, Trimer & 1.1264      & --0.4758$^*$  \\
\hline
Atom                   & $\epsilon$  & $\sigma$     \\
\hline 
Mg                     & 0.88        &    1.64      \\
Cl                     & 0.71        &    4.02      \\
\hline\hline
\end{tabular*}
{\small$^*$Compensating Cl$^-$ counterions were assigned a charge of --1
to maintain charge neutrality.}
\end{table} 

\noindent The GAFF force field parameters for THF were benchmarked
against the experimental properties and found to reproduce the
experimental values adequately; for example the experimental density of
THF ($\sim$ 0.889~g/cm$^3$) is well reproduced by CMD simulations
($\sim$ 0.882~g/cm$^3$),\cite{Rajput2015} similarly the experimental
diffusion coefficient ($\sim$ 300.00$\times$10$^{-11}$~m$^{2}$/s) is in
good agreement with the calculated value ($\sim$
211.34$\times$10$^{-11}$~m$^{2}$/s).\cite{Rajput2015}

The MACC electrolyte structures  initially optimized with Gaussian09
(see above) are inserted into a periodic box of size
48$\times$48$\times$48~\AA$^3$ containing 800 THF molecules at the
experimental THF density (0.889 g/cm$^3$). The infinite dilution limit
is simulated for each complex, i.e.\ only one molecule was inserted in
the CMD box. Then, each configuration is equilibrated for 1~ns in the
isothermal-isobaric ensemble (NPT) which is sufficient to converge the
density, with minimal variation ($\sim$ 1\% ) from the THF experimental
value. Subsequently, a 1~ns simulation is performed in the canonical
ensemble (NVT) at 300~K, of which the first 200~ps is utilized for
equilibration, within which convergence of each simulation is achieved,
followed by a production time of 800~ps. A time step of 1~fs is used. 

To identify how THF coordinates to the Mg$_x$Cl$_y$ (i.e.\ monomer,
dimer and trimer) ionic species, it is not necessary to consider changes
of the ion structures during coordination by THF. Therefore, in the CMD
simulations the ions are held rigid (in the electrolyte solvent) at the
fully relaxed geometries as obtained by Gaussian09, thus removing the
necessity of parameterizing bonded interactions of each ion.


\section{Results}
\label{sec:results}

To isolate the electro-active species comprising the MACC electrolyte,
we first study the structures and composition of various magnesium
chloride complexes hypothesized to be present in the electrolyte. The
Mg-Al-Cl-THF chemical space is further enlarged by additional structures
that are guessed by chemical intuition or results of CMD simulations.
Consequently, we study the salt solvation by altering the first
solvation shell of the magnesium-chloro complexes considered. The
Supplementary Information reports  the atomic positions of the
thermodynamic stable structures.

\subsection{Magnesium-chloride complexes}
\label{sec:method}

Previous experimental efforts have attempted to understand the complex
structure of the magnesium-chloride complexes of the MACC electrolyte.
The combined X-ray diffraction, Raman, and NMR spectroscopies by Aurbach
and co-workers established that Mg$^{2+}$ in THF exists always as a six
coordinated ion in the form of monomer MgCl$^+$(5THF)  or dimer
Mg$_2$Cl$_3^+$(6THF).\cite{Gizbar2004,Pour2011} In contrast to these
results, a more recent theoretical investigation\cite{Wan2014}
elucidated the first solvation shell of the MgCl$^+$ and Mg$_2$Cl$_3^+$
magnesium organo-chloro species in the bulk electrolyte using \emph{ab
inito} molecular dynamics calculations, and suggested that the MgCl$^+$
monomer is always coordinated by three THFs, leading to a total Mg
coordination of four. Similar findings were supported by the
experimental NMR and XANES work of Nakayama \emph{et
al}.\cite{Nakayama2008}

We benchmark our modelling strategy on previous experimental and
theoretical results by simulating several magnesium-chloride complexes
in different THF environments combining DFT and CMD calculations, as
outlined above. 

To measure the stability of magnesium-chloride clusters in THF we
compute (with DFT) the formation free energy $\Delta F$ at fixed THF
chemical potential $\mu _{\rm THF}$:
\begin{equation}
\label{eq:formationgibbs}
\Delta F = G(n_{\rm Mg}, n_{\rm Cl}, n_{\rm THF}) - G(n_{\rm Mg}, n_{\rm Cl}) -n_{\rm THF}\mu_{\rm THF}
\end{equation}
where $G(n_{\rm Mg}, n_{\rm Cl})$ and $G(n_{\rm Mg}, n_{\rm Cl}, n_{\rm
THF})$ of Eq.~\ref{eq:formationgibbs} are the Gibbs free energy of each
Mg$_x$Cl$_y$ cluster isolated and coordinated by $n_{THF}$ molecules.
Throughout the paper all references to ``formation~energy'' refer to the
formation free energy $\Delta F$. 

\begin{figure}[!t]
\centering
\includegraphics[width=1.0\columnwidth]{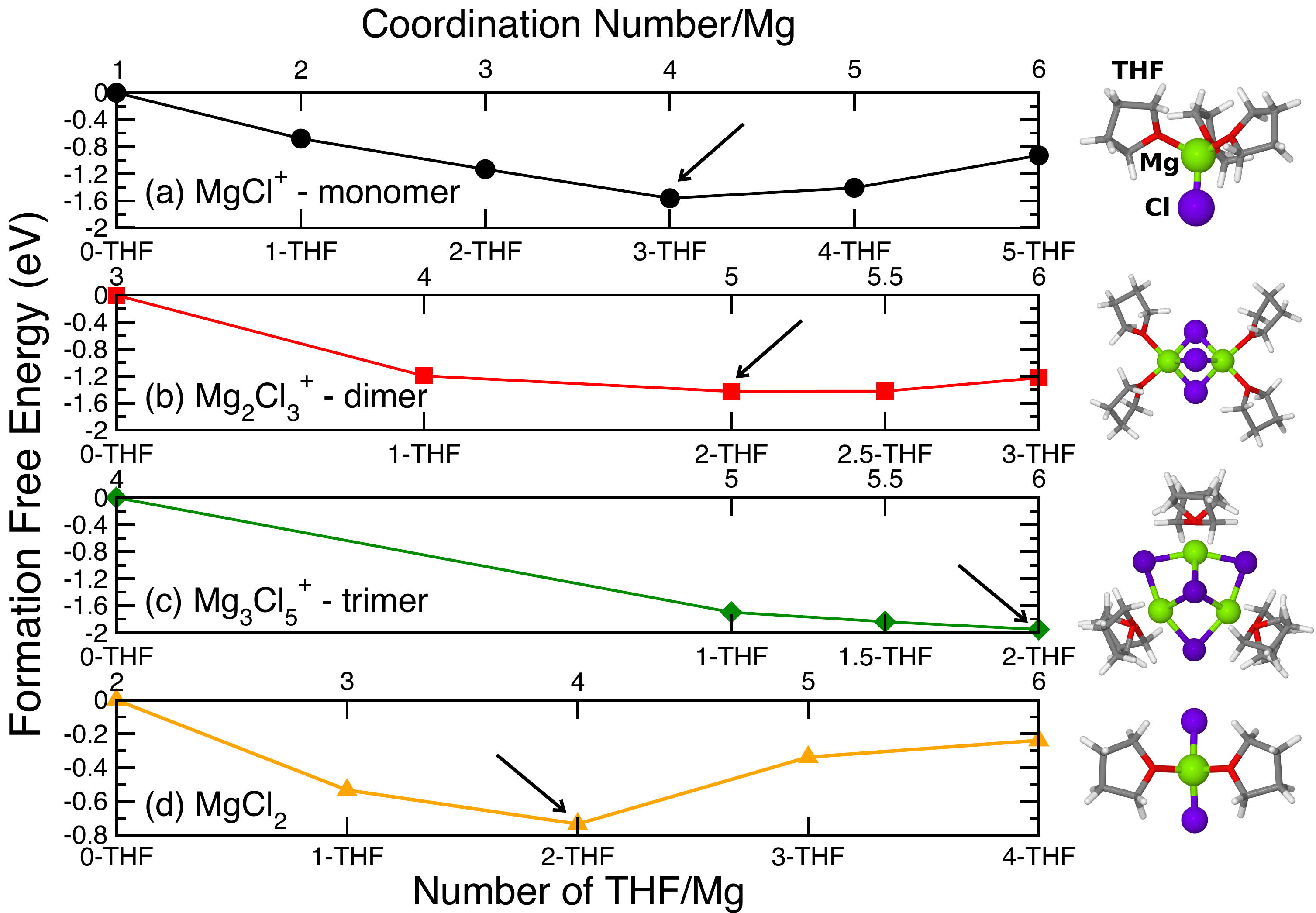}
\caption{\label{fig:complex_hulls} Formation free energy (in eV) of
magnesium-chloride complexes as a function of THF coordination for a)
MgCl$^+$ (monomer), b) Mg$_2$Cl$_3^+$ (dimer), c) Mg$_3$Cl$_5^+$
(trimer), and d) MgCl$_2$. Arrows indicates the most stable THF
coordination environment for each complex.  Snapshots of the most stable
magnesium-chloride complexes are also depicted.}
\end{figure}

Figure~\ref{fig:complex_hulls} depicts the free energies of formation
$\Delta F$ for the magnesium-chloride complexes as a function of THF
coordination (bottom $x$-axis) and total Mg coordination (top $x$-axis)
obtained from B3LYP calculations. 

Each minimum in Fig.~\ref{fig:complex_hulls} represents the most stable
structure for a particular Mg-Cl complex, hence its most stable Mg
coordination. Figure~\ref{fig:complex_hulls} shows that the preferred
magnesium coordination is 4-fold for both MgCl$^+$(3THF) and
MgCl$_2$(2THF), 5-fold for the dimer Mg$_2$Cl$_3^+$(4THF), and 6-fold
for the trimer Mg$_3$Cl$_5^+$(6THF). Interestingly, the total Mg
coordination number increases with the size of the magnesium-chloride
cluster. The stable structures show a coordination of 3 THFs for the
monomer MgCl$^+$ and 2 THFs for each Mg atom in dimer Mg$_2$Cl$_3^+$
(4THFs in total). These results are consistent with theoretical findings
by Wan \emph{et al.}~\cite{Wan2014} and  XANES spectroscopy
data.\cite{Nakayama2008} 

\begin{figure}[h]
\centering
\includegraphics[width=1.0\columnwidth]{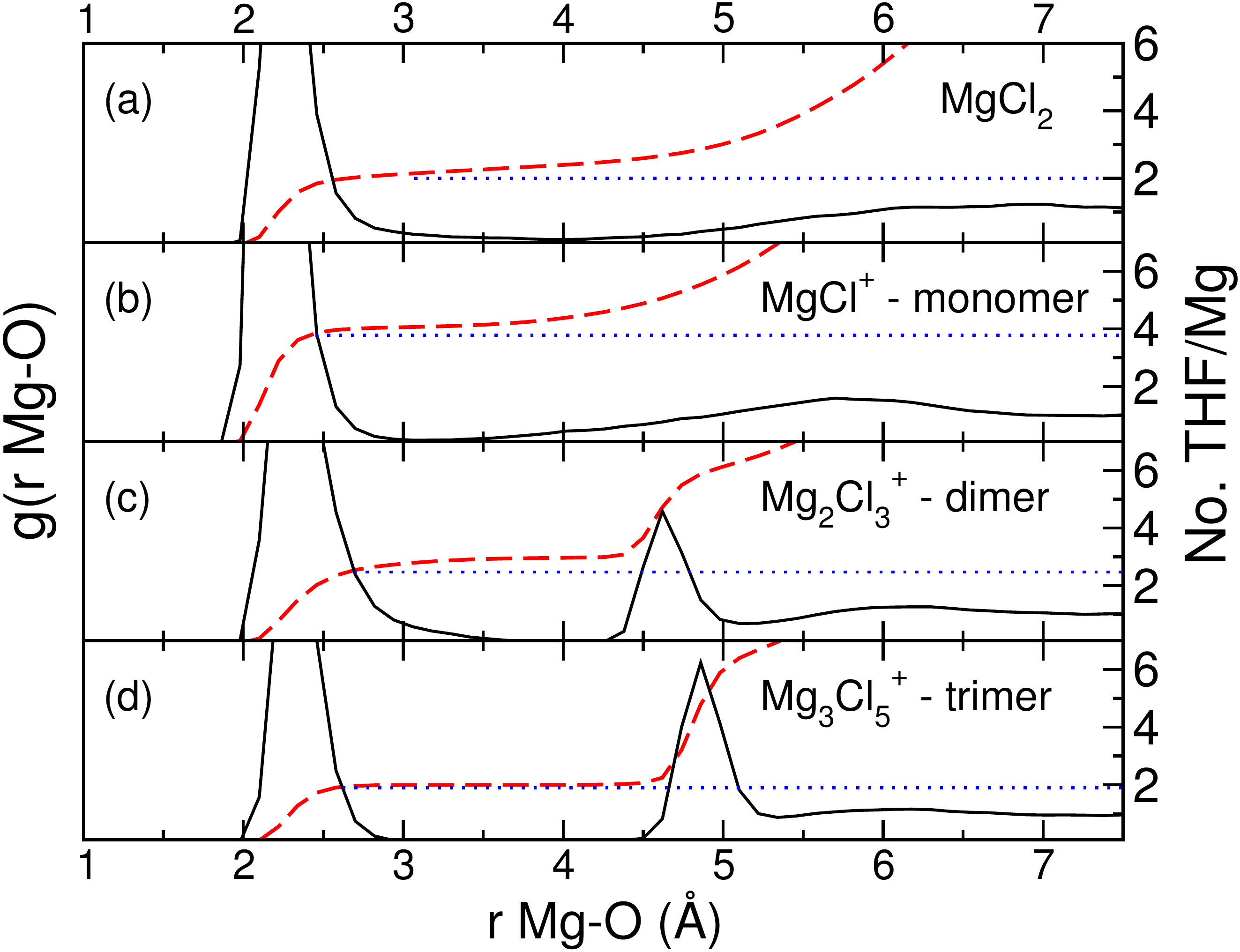}
\caption{\label{fig:md_rdf}Mg-O(THF) radial distribution function (black
curves), and coordination numbers (red dashed lines) as function of the
Mg-O separation (in \AA) for the magnesium chloride-complexes. a)
MgCl$_2$, b) MgCl$^+$ monomer, c) Mg$_2$Cl$_3^+$ dimer, and d)
Mg$_3$Cl$_5^+$ trimer.  The THF coordination numbers for the first
coordination shell are indicated by the blue dotted lines.} 
\end{figure}

Classical molecular dynamic simulations are used to clarify the dynamics
of the ion complexes in MACC in THF solvent.  Figure~\ref{fig:md_rdf}
plots the radial distribution functions (RDF, black lines), and the
corresponding coordination numbers for the 4 complexes (red and blue
dashed and dotted lines, respectively) obtained from CMD calculations.
More RDF plots of different atom pairs are available in the
Supplementary Information.

The peaks located between 2.0-2.5 \AA{} in each RDF indicate the first
coordination shell of Mg experienced by THF, and its integration (see
blue dotted lines) the number of THF molecules coordinated by Mg atoms.
For dimer and trimer of Fig.~\ref{fig:md_rdf}~c)~and~d) the peaks from
4.5 to 5.0~\AA{} are the oxygen atoms of the THFs coordinating  the
nearest neighbors  Mg atoms.  Figure~\ref{fig:md_rdf} obtained from
classical molecular dynamic simulations, shows that the overall Mg
coordination number for each Mg-Cl complex is consistent with the
prediction from the DFT formation energies (see above). Our preliminary
CMD simulations demonstrated that the coordination of MgCl$^+$ is
wrongly predicted to be 6, if Mg$^{2+}$ and Cl$^-$ ionic charges are
assigned to +2 and --1 for Mg and Cl ions,
respectively.\cite{Wan2014,Nakayama2008}  This is because charge
transfer processes occurring within each complex reduce the nominal
charges on both Cl and Mg (as seen in Table~\ref{tb:charges}), hence
lowering the overall Mg coordination number to 4. Moreover, we find two
kinds of Mg-THF coordinations observed for the charged ions-- THF
molecules are either strongly coordinated to the complex, or weakly
coordinated, thereby setting up a free exchange of THF molecules with
the bulk solution and the first solvation shell. For Mg$_2$Cl$_3^+$, two
THF molecules are strongly coordinated to each Mg atom, while a third
THF molecule is constantly exchanged between the bulk region and the Mg
atoms in an alternating manner, leading to an effective coordination
number of 2.5 THFs per Mg atom. For instance, this exchange occurs after
900 ps for Mg$_2$Cl$_3^+$. For the MgCl$^+$ monomer, the coordination
number is predicted to be 3.7 THFs, and slightly larger than in previous
\emph{ab initio} MD simulations (3 THFs for MgCl$^+$).\cite{Wan2014}

Overall, the coordination numbers computed from both DFT and CMD are
consistent with previous XANES,\cite{Nakayama2008} with sub-ambient
pressure ionization with nano-electrospray mass
spectroscopy,{\cite{Liu2015a}} and accurate \emph{ab initio} MD studies
consolidating the idea that THF steric hindrances and Mg-Cl charge
transfer lead to THF not being able to fulfill the typical sixfold
coordination of Mg in solids.\cite{Wan2014} To conclude, DFT coupled
with computationally inexpensive CMD simulations provides a robust
strategy to interrogate the structural characteristic of the species in
the MACC electrolyte.

To study the effect of electrolyte composition on the stability of the
MACC complexes, we analyze the stable phases of the Mg-Cl-Al-THF
chemical space using the the total energies of more than hundred
Mg$_x$Cl$_y$ and Al$_x$Cl$_y$ molecules with variable THF coordination
numbers. 

\begin{figure}[h!]
\centering
\includegraphics[width=1.0\columnwidth]{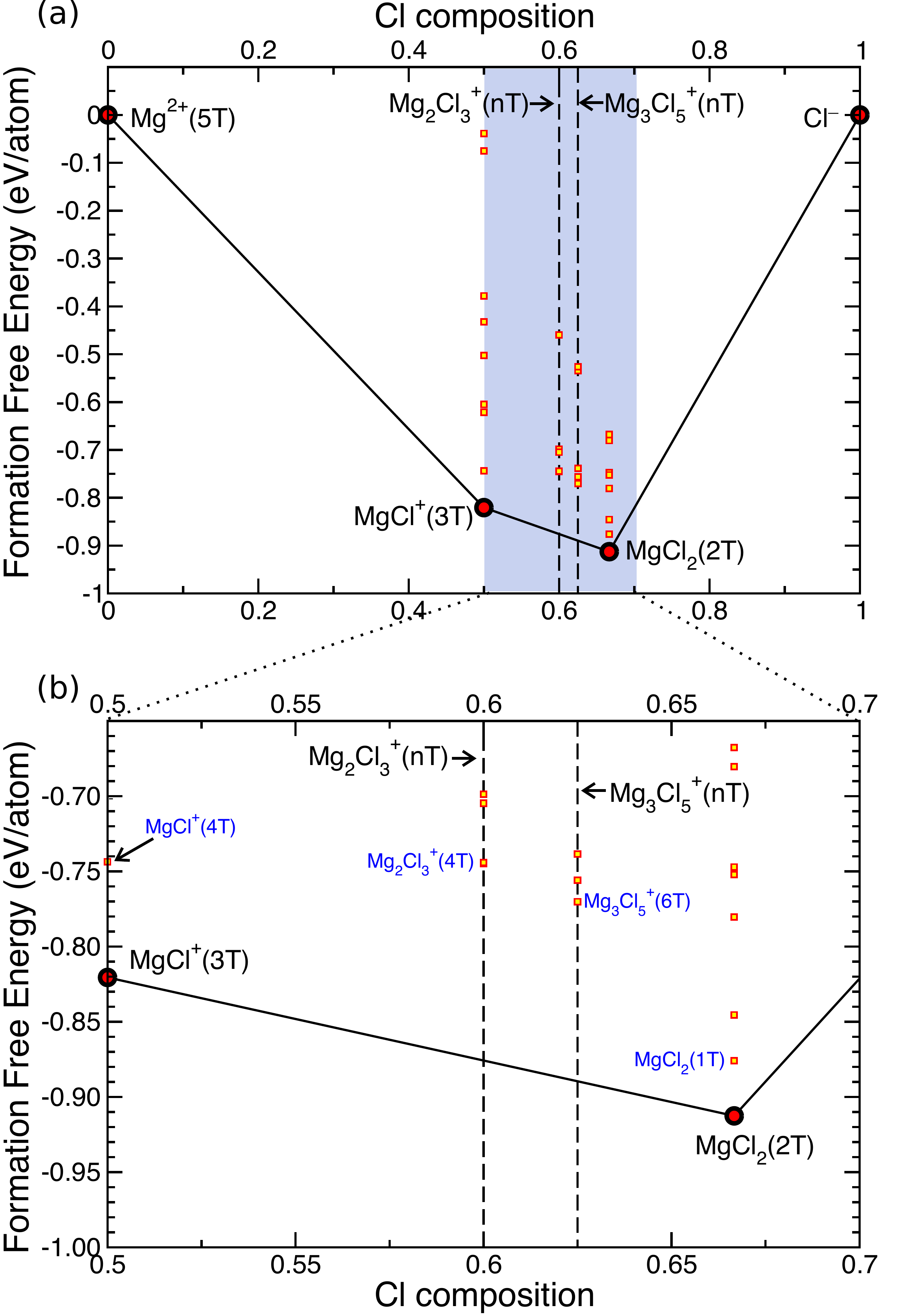}
\caption{\label{fig:mg_cl_thf_grandphasediag} a) Mg-Cl-THF
grand-potential phase diagram as function of Cl concentration at the
bulk THF chemical potential. Red dots connected by black lines indicate
the stable magnesium-chloride complexes, and unstable species by yellow
squares. b) shows a region close to the stability lines. Coordinating
THF molecules indicated by T. Dashed lines for the dimer and trimer
concentrations. }
\end{figure}

Figure~\ref{fig:mg_cl_thf_grandphasediag} shows the grand-potential
phase diagram\cite{Ong2008} for the Mg-Cl-THF system at the bulk THF
chemical potential, where black lines set the boundaries of the stable
regions, and red-dots indicate the stable phases.  To the best of our
knowledge this is the first instance of grand-potential phase diagrams
together with DFT calculations being used to analyze the structure of a
liquid electrolyte. 

The THF chemical potential is calculated using the same procedure
exposed in Sec.~\ref{sec:method} from a cluster of 7 THF molecules
extracted from an well equilibrated \emph{ab initio} MD, see more
details in Ref.~\cite{Canepa2015} The grand potential formation free
energy of each complex is calculated with respect to Mg$^{2+}$ in THF
and Cl$^-$ in THF as these are the relevant reference states for the
complexes in THF.  At Cl compositions of 0.0 (equivalent to 0 \% Cl and
100 \% Mg, or Mg$^{2+}$) and 1.0 (equivalent to 100 \% Cl and 0 \% Mg,
or Cl$^-$), Mg$^{2+}$ and Cl$^-$ species are coordinated  by 6 and 0
THFs, respectively.  The formation free energies of Fig.{~\ref{fig:mg_cl_thf_grandphasediag}} and successive grand-potential phase diagrams contain the vibrational entropy  as indicated by Eq.~{\ref{eq:foamtionenergy}} and Eq.~{\ref{eq:formationgibbs}.  Our calculations do not capture the configurational entropy that would tend to stabilize low coordination number.}
At first glance
Fig.~\ref{fig:mg_cl_thf_grandphasediag}a and b show that solutions
containing Mg and Cl in THF form stable magnesium-chloride complexes,
and only MgCl$^+$(3T) (with T for THF) and MgCl$_2$(2T) are observed to
be the stable phases (see red dots) through the entire Cl composition.
We find that neither the dimer, nor the trimer are stable in bulk THF
(see yellow squares marked as Mg$_2$Cl$_3^+$(4T) and Mg$_3$Cl$_5^+$(6T)
for the dimer, trimer, respectively in
Fig.~\ref{fig:mg_cl_thf_grandphasediag}b).  Notably, the stable phases
identified by the grand-potential phase diagram of
Fig.~\ref{fig:mg_cl_thf_grandphasediag} correspond to the lowest
formation energies of the magnesium-chloride complexes as presented in
Fig.~\ref{fig:complex_hulls}. As already mentioned, the dimer
Mg$_2$Cl$_3^+$ was isolated with X-ray on mono-crystals as one of the
products of crystallization of the APC electrolyte,\cite{Pour2011} while
trimer and higher order structures were speculated to exist by
Barile~\emph{et al}.\cite{Barile2014} as a byproduct of polymerization
of the principal MACC components MgCl$_2$, monomer and dimer.

\begin{figure}[h!]
\centering
\includegraphics[width=1.0\columnwidth]{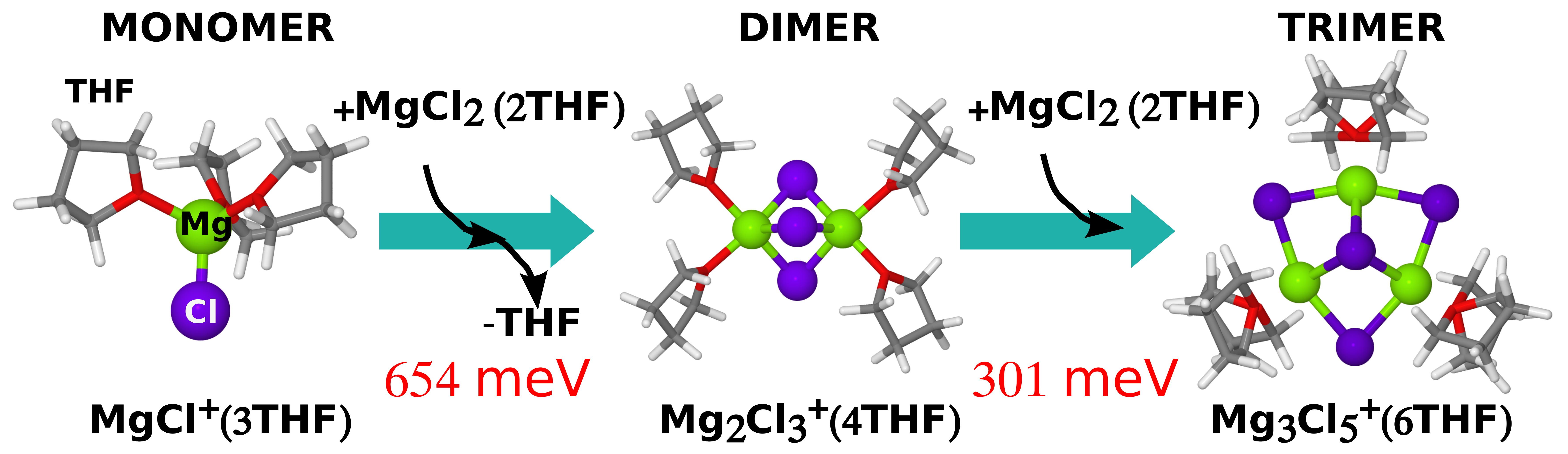}
\caption{\label{fig:reaction} Chemical reaction for monomer converting
into dimer and trimer, with relative free energy (in meV). }
\end{figure}

By isolating the most stable Mg$_x$Cl$_y$ components from the
grand-potential phase diagram of Fig.~\ref{fig:mg_cl_thf_grandphasediag}
we can estimate the reaction energies to form dimer and trimer complexes
from the MgCl$^+$ as seen in Fig.~\ref{fig:reaction}. 
Figure~\ref{fig:reaction} clearly shows that both dimer and trimer are
not accessible in normal thermodynamic conditions (with almost 1~eV to
create the trimer from the monomer). Hence our results argue against
the existence of such magnesium-chloride agglomerates in the actual
solution.\cite{Pour2011,Benzmayza2013,Doe2014,Barile2014}

To explore  the morphology of the MACC electrolyte further and analyze
possible changes that might occur to its structure, we investigate the
stability of the electrolyte with varying THF chemical potential, and
therefore look for possible conditions under which unstable structures
from Fig.~\ref{fig:mg_cl_thf_grandphasediag} can be stabilized (e.g.\
dimer and trimer).

\begin{figure}[!ht]
\centering
\includegraphics[width=1.0\columnwidth]{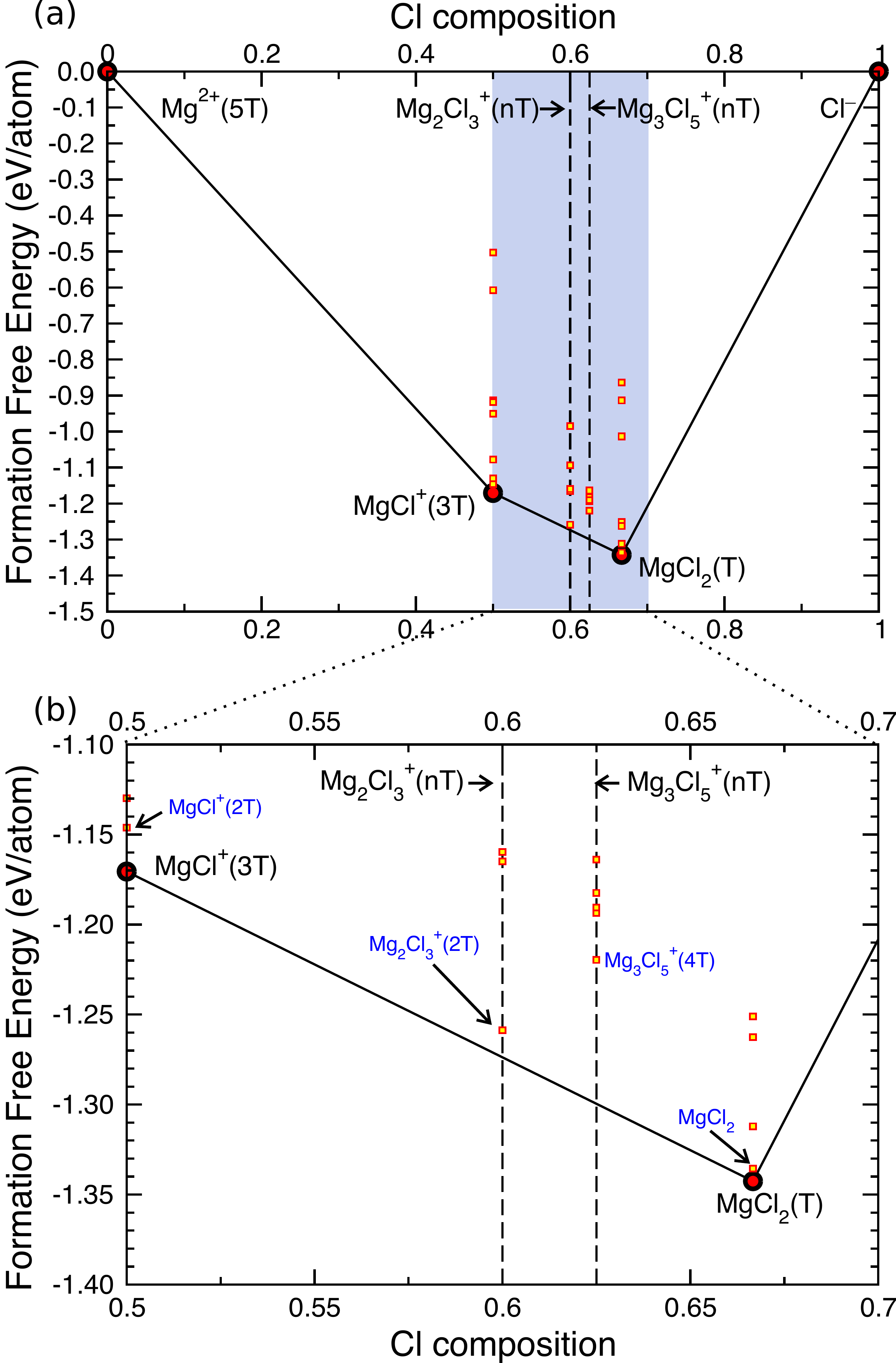}
\caption{\label{fig:mg_cl_thf_grand_drying} a) Mg-Cl-THF grand-potential
phase diagram as function of Cl concentration at the THF chemical
potential under drying conditions (with activity of THF $=$ 10$^{-6}$).
Red dots connected by black lines indicate the stable magnesium chloride
complexes, and unstable species are indicated by yellow squares. b)
shows a region close to the stability lines. Coordinating THF molecules
indicated by T. Dashed lines as guide for dimer and trimer
concentrations.  }
\end{figure}

Figure~\ref{fig:mg_cl_thf_grand_drying} shows the grand-potential phase
diagram\cite{Ong2008} for the Mg-Cl-THF system at the chemical potential
corresponding to a THF activity of $10^{-6}$ (which means lowering the
THF chemical potential by $\sim$~--34.25~kJ~mol$^{-1}$).
Changing the THF chemical potential affects directly the relative
stability of the THF coordination environment experienced by the MACC
complexes, and low-coordination situations are preferred at low THF
chemical potentials. For example, the THF coordination of MgCl$_2$
decreases from 2THF to 1THF, but the relative stability order between
MgCl$_2$ and MgCl$^+$ is maintained. Lowering the chemical potential of
THF in the bulk solution emulates the process of drying, where the
solvent evaporates, and thereby only the strongly-bound THF molecules
remain coordinated to the complexes. 

Interestingly, from Fig.~\ref{fig:mg_cl_thf_grand_drying} we notice that
the free energy of the dimer complex now approaches the ground state
line (see black line), meaning that this structure might become
accessible under conditions of evaporating/drying solvent. The dimer is
only $\sim$~0.02 eV above the stability line that is enough to be
accessible by thermal fluctuations, and may explain why the dimer was
successfully crystallized.\cite{Pour2011}  In summary our results show
that in pure Mg-Cl-Al-THF solutions neither the dimer nor the trimer
exist at equilibrium, but a reduction of the THF chemical potential, as
experienced in drying, may lead to the formation of dimers.

\subsection{Aluminum complexes in THF and Al-Cl-Mg Phase Diagram}

\begin{figure}[t]
\centering
\includegraphics[width=1.0\columnwidth]{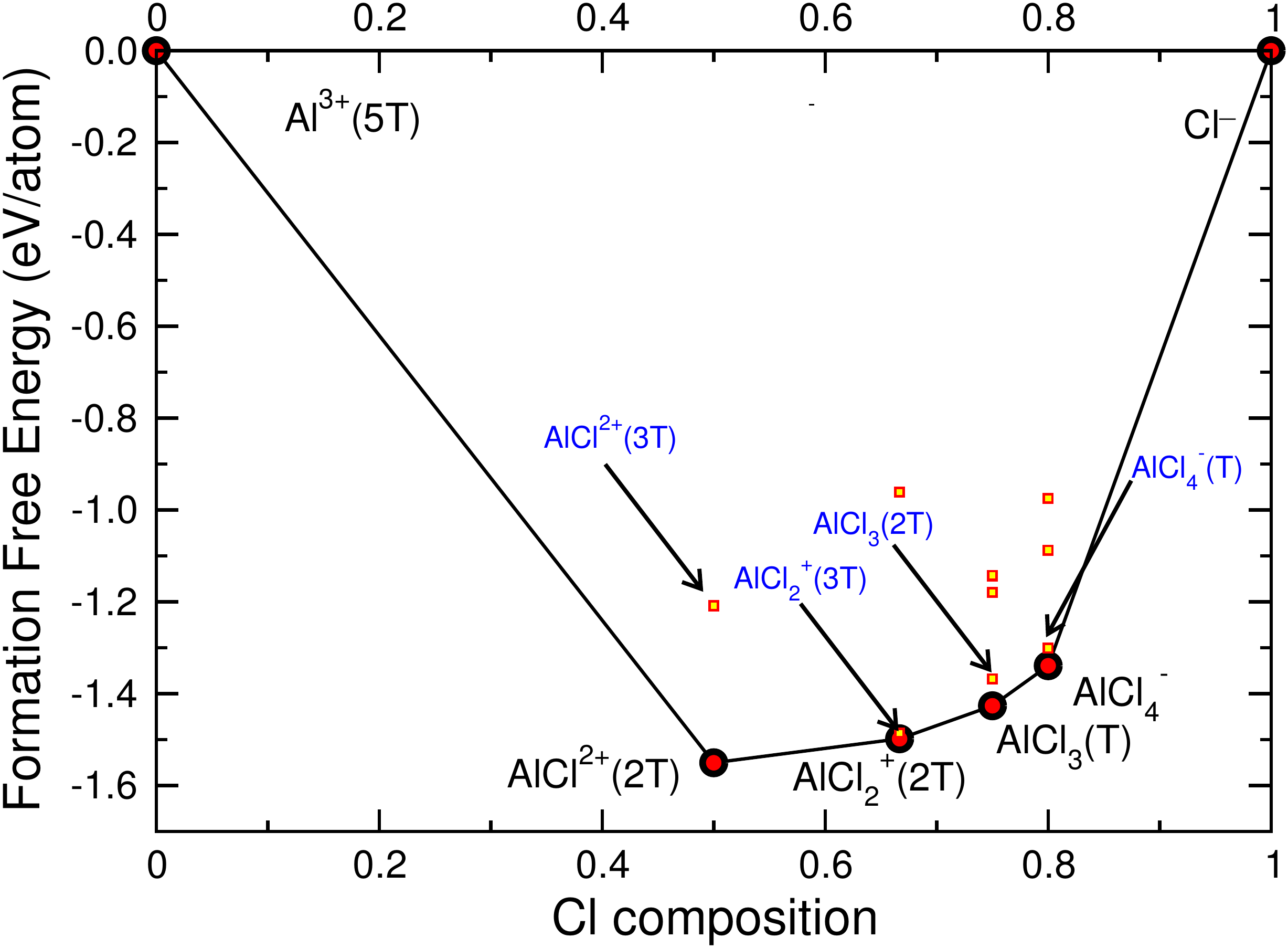}
\caption{\label{fig:Al_Cl_THF_grandphasediag} Al-Cl-THF grand-potential
phase diagram as function of Cl concentration at the THF chemical
potential. Red dots connected by black lines indicate the stable
magnesium-chloride complexes, and unstable species are indicated by
yellow squares.  Coordinating THF molecules are indicated by T.}
\end{figure}

To complete the thermodynamic analysis of the possible species in the
MACC electrolyte we perform a similar study of Al-Cl in THF.
Figure~\ref{fig:Al_Cl_THF_grandphasediag} shows the Al-Cl-THF
grand-potential phase-diagram, where at Cl compositions of 0.0 and 1.0
are located the isolated species of Al$^{3+}$ and Cl$^-$ coordinated by
5 and by 0 THFs, respectively.

Figure~{\ref{fig:Al_Cl_THF_grandphasediag}} suggests that the stable
species in the electrolyte are: AlCl$_2^{2+}$(2THF), AlCl$_2^+$(2THF),
AlCl$_3$(THF) and  AlCl$_4^-$, with no Al$_x$Cl$_y$ polymeric species
found to be stable in THF. Figure~\ref{fig:Al_Cl_THF_grandphasediag}
also captures the coordination of the aluminum chloride species by THF,
with fourfold coordination for AlCl$_3$(THF); AlCl$_4^-$ being already
fourfold coordinated does not have strongly bonded THF molecules. 

\begin{figure}[t]
\centering
\includegraphics[width=0.9\columnwidth]{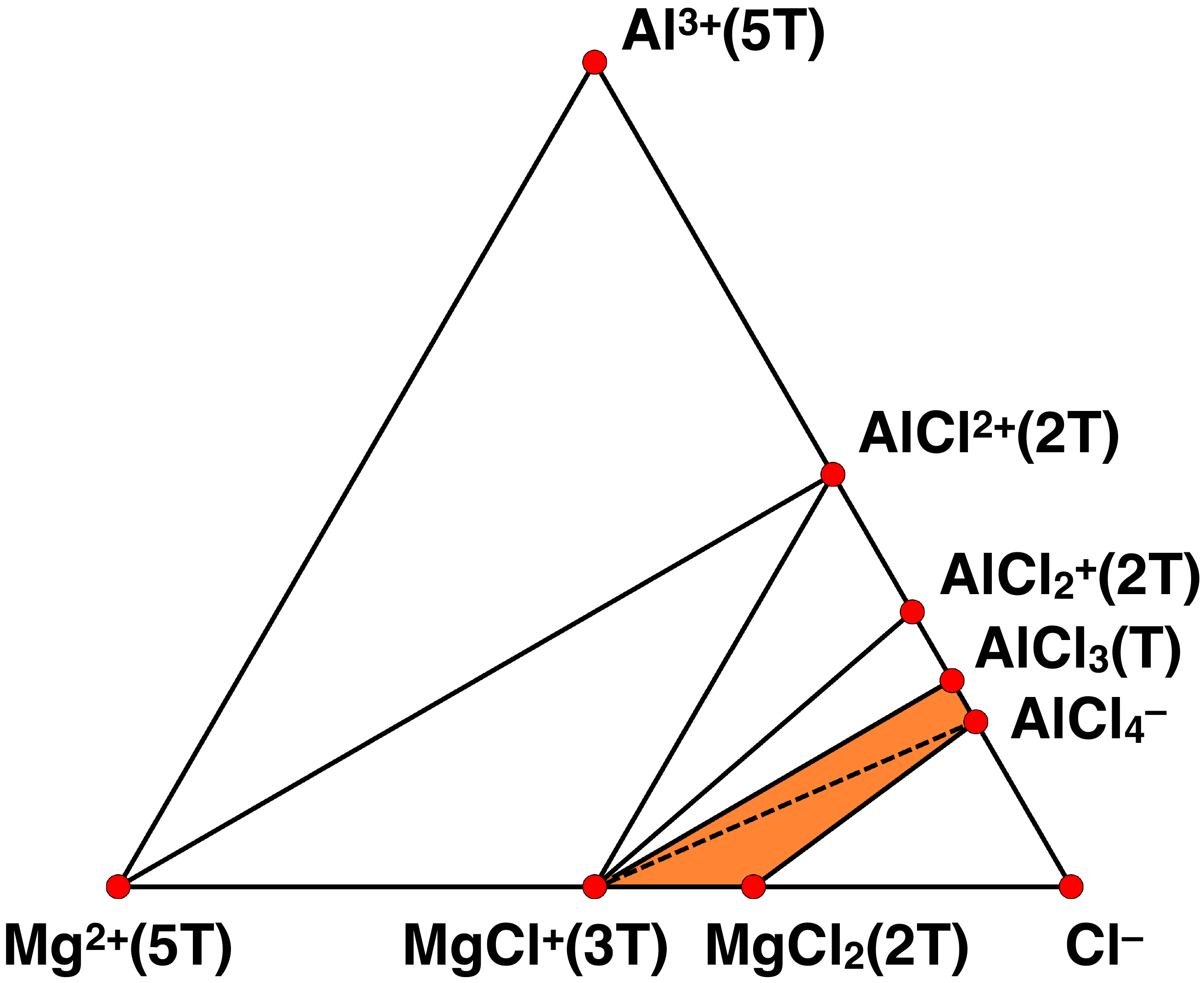}
\caption{\label{fig:Al_Cl_Mg-THF_pd} Liquid Al-Cl-Mg-THF
grand-potential phase diagram at the THF chemical potential. Red dots
connected by black lines indicate the stable magnesium- and
aluminum-chloride complexes. Coordinating THF molecules are indicated by
T.  Orange area and dashed line indicate the exchange reaction (c) (see
Table~{\ref{tb:equil_reactions}} and text for more details). Tie-lines do not necessary  respect charge neutrality (see text and Supplementary Information).}
\end{figure}

Figure~{\ref{fig:Al_Cl_Mg-THF_pd}} shows the ternary Al-Cl-Mg
grand-potential phase-diagram  at the THF chemical potential and black
lines indicate tie-lines (more detail in the Supplementary Information). Although the ternary Al-Cl-Mg grand-potential phase-diagram shows
tie-lines such as  Mg$^{2+}$ -- AlCl$_2^+$(2THF), MgCl$^+$(3THF) -- AlCl$_2^+$(2THF), and MgCl$^+$(3THF) -- AlCl$^{2+}$(2THF)   these species cannot
co-exist as  they do not respect the charge neutrality of the MACC
electrolyte. The orange part of the PD  in
Fig.~{\ref{fig:Al_Cl_Mg-THF_pd}}  represents the zone  where ionic
species can co-exist and respect charge neutrality.   Therefore, from the
Al-Cl-THF grand-potential PD we deduce that the important equlibria will
only occur among MgCl$^+$, AlCl$_4^-$ and MgCl$_2$ and AlCl$_3$ species
and the dashed tie-line indicate their interaction.  In the orange area, the only plausible equilibria that  respect charge neutrality are either MgCl$^+$(3THF) with AlCl$_4^-$ or MgCl$_2$(2THF) with AlCl$_3$(THF). We use a dashed line
for the interaction between MgCl$^+$(3THF) and AlCl$_4^-$ to indicate
the small driving force of this reaction (see $\Delta$E$_{D-H}$ of
reaction (c) in Table~{\ref{tb:equil_reactions}}), which will be
discussed in more detail later. The stable composition of the MACC electrolyte is the  MgCl$^+$(3THF) and AlCl$_4^-$ tie-line.

Despite the small dielectric constant of THF ($\sim$~7.58 at 298~K),
from the phase diagram of Figure~{\ref{fig:Al_Cl_Mg-THF_pd}} we do not
find stable ionic couples (e.g.\ MgCl$^+$AlCl$_4^-$) that would hinder
the electrochemical function of the electrolyte. For example, the
formation free energy of the monomer and the dimer ionic couples
[MgCl$\cdot$(3THF)]$^+$AlCl$_4^-$  and
[Mg$_2$Cl$_3$$\cdot$(4THF)]$^+$AlCl$_4^-$ require $\sim$~0.064~eV and
$\sim$~0.088~eV, respectively. 


\subsection{MACC under equilibrium and conditioning}
We use the knowledge of the stable species gained from grand-potential
phase diagrams to explain the phenomenological effects observed in the
MACC electrolyte under electrochemical cycling. 

MACC electrolytes exhibit high coulombic efficiency, but only  after
extensive electrochemical cycling, a process termed
conditioning.\cite{Shterenberg2014}  Barile \emph{et
al.}\cite{Barile2014} demonstrated that when a conditioned MACC
electrolyte is left to rest for a prolonged period of time it shows
lower coulombic efficiencies than when conditioned, and referred as
``aging'' of the electrolyte.  The changes in the electrolyte species
caused by aging and conditioning can be rationalized by evaluating
possible reaction equilibria occurring in bulk and at the electrodes
summarized in Table~{\ref{tb:equil_reactions}}.  The grand-potential
phase diagrams in Fig.~\ref{fig:mg_cl_thf_grandphasediag},
Fig.~\ref{fig:mg_cl_thf_grand_drying},
Fig.~\ref{fig:Al_Cl_THF_grandphasediag} and
Fig.~\ref{fig:Al_Cl_Mg-THF_pd} attest to the presence of only four
stable magnesium/aluminum chloride species, MgCl$_2$, MgCl$^+$,
AlCl$_3$ and AlCl$_4 ^-$ limiting the total number of species in the
reactions of Table~\ref{tb:equil_reactions}.   Here, we do not consider
the ionic dissociation of  MgCl$_2$ (ALCl$_3$) in Mg$^{2+}$ (Al$^{3+}$)
and Cl$^-$ because of their high-energy in our simulations.  In
addition, DFT calculations and \emph{ab initio} MD dynamics confirm that
Cl$^-$ is poorly coordinated by THF.

\begin{table*}[!ht]
\caption{\label{tb:equil_reactions}Possible reaction equilibria of the
Al-Cl-Mg-THF system, $\Delta$E and corrected by Debye-H\"{u}ckel
$\Delta$E$_{D-H}$ (in eV). $\Delta$E are computed from the total energy ($E_{\rm PCM}$ for liquid molecules) of each species.
}
\begin{tabular*}{\textwidth}{@{\extracolsep{\fill}}lcccr@{}}
\hline \hline
\multicolumn{2}{c}{Reaction}  &  $\Delta$E            &                             $\Delta$E$_{D-H}$                                           \\        
\hline
(a) & MgCl$_2$(s) + 2THF(l) $\leftrightarrow$ MgCl$_2$(2THF)(l)         &  0.251    &     --  \\
(b) & AlCl$_3$(s) + THF(l) $\rightarrow$  AlCl$_3$(THF)(l)              & --1.138   &     --  \\ 
(c) & MgCl$^+$(3THF)(l) + AlCl$_4^-$(l) $\leftrightarrow$  MgCl$_2$(2THF)(l) + AlCl$_3$(THF)(l)                & --0.106    &     0.085   \\ 
\hline
 \multicolumn{2}{c}{Mg displacement of Al} &$\Delta$E & $\Delta E_{D-H}      $\\
\hline
(d) & 2AlCl$_3$(THF)(l) + 3Mg(s)  + 4THF(l)  $\rightarrow$   3MgCl$_2$(2THF)(l) + 2Al(s)                                     & --2.186     &  --         \\    
(e) & AlCl$_4^-$(l) + MgCl$^+$(3THF)(l) + 1.5Mg(s) + 2THF(l) $\rightarrow$ 2.5MgCl$_2$(2THF)(l) + Al(s)                      & --1.199     & --1.033    \\ 
\hline \hline
\end{tabular*}
\end{table*}

The reaction energy  ($\Delta$E) at the dilute limit, and with the
Debye-H\"{u}ckel correction ($\Delta$E$_{D-H}$) are included in
Table~{\ref{tb:equil_reactions}}.  The  Debye-H\"{u}ckel  correction
captures the electrostatic interaction of charged species in solution at
dilute activities, and stabilizes the  ions in the electrolyte, thus
affecting some reaction energies of Table~{\ref{tb:equil_reactions}}. By
fixing the MACC concentrations at the typical experimental value of
0.5~M,{\cite{Doe2014}} the computed $\Delta$E$_{D-H}$ correction for
reactions (c) and (e) is substantial and $\sim$~0.1914 eV.

Reaction (a) of Table~\ref{tb:equil_reactions} dictates the equilibrium
of MgCl$_2$ between its liquid and solid state, a reaction which is
predicted as endothermic. The magnitude of the $\Delta E$ shows that
MgCl$_2$ is  sparingly ``dissolved'' in ethereal organic solvents such
as THF or glymes and is supported by  previous experimental
evidences.\cite{Doe2014,Liao2015} Reaction (b) that sets the
``dissolution'' of AlCl$_3$ in THF, is highly exothermic suggesting that
AlCl$_3$ occurs in liquid THF. 

In order to maintain charge neutrality, the activities of the charged
species in solution, namely MgCl$^+$(3THF)(l) and AlCl$_4^-$(l), must
remain equal, and this condition is regulated by reaction (c) of
Table~\ref{tb:equil_reactions}. Reaction (c) is slightly exothermic,
favoring the formation of neutral molecules (MgCl$_2$ and AlCl$_3$) in
the electrolyte. Nevertheless, when the Debye-H{\"{u}}ckel
correction is applied to reaction (c), the formation of ions is favored
guaranteeing the operability of the MACC electrolyte. This stresses the
importance to include the effect of the ion activities to compute
properly reaction energies in liquids. Moreover, for the ionic strengths
in the MACC electrolyte, the Debye-H{\"{u}}ckel correction is
sufficient, and no higher order corrections are necessary. The
conductivity of the MACC electrolyte is related directly to the
concentration of the charged ionic species MgCl$^+$, and AlCl$_4^-$. The
slightly endothermic nature of reactions (c) $\Delta$E$_{D-H}$ shows
that under thermodynamic equilibrium the charged species MgCl$^+$ and
AlCl$_4^-$ are present in the electrolyte.  At the solubility limit
of MgCl$_2$ $\sim 7.8\times10^{-4}$ M in THF (set by reaction (a)) and for a 0.5~M of AlCl$_3$ in
THF,  the MgCl$^+$ activity  is approximately 92~mM, which is high
enough to guarantee good ionic  conductivity (see discussion later). Note that the Debye-H{\"{u}}ckel  correction on the $\Delta$E of reaction (c) is concentration dependent (see Eq.{~\ref{eq:screening_length}}) and is calculated for a 0.5 M solution of  MACC electrolyte.

The processes of non-electrochemical Mg and Al deposition are regulated
by reactions (d) and (e) in Table~\ref{tb:equil_reactions}. Reaction (d)
depicts the equilibrium between magnesium aluminum chloride neutral
species and Mg and Al metals. The highly exothermic character of
reaction (d) explains that Al deposition is preferred at the cost of Mg
dissolution.  A similar trend is observed for reaction (e) that
establishes the equilibrium of charged and neutral magnesium aluminum
chloride species and the respective metals. The reduction potential of
Al ($\sim$~--1.67~V vs.\ NHE) is more positive than for Mg
($\sim$~--2.35~V vs.\ NHE) and ensures immediate Al deposition during
initial electrochemical cycles. Spontaneous Al deposition sets a
thermodynamic driving force for the process of aging, in absence of an
applied potential at the electrode. 

According to reactions (d) and (e)  of Table~\ref{tb:equil_reactions}
when a conditioned electrolyte is allowed to rest (i.e.\ not undergoing
electrochemical cycling) the concentration of the electroactive species
available in solution, MgCl$^+$ and AlCl$_4^-$, decrease by several
orders of magnitude as Al ions in solutions are deposited on the
electrode. Though the contribution of Debye-H\"{u}ckel correction is
substantial on the $\Delta$E of reaction (e), it is not sufficient to
stop Al deposition. We speculate that the spontaneous nature of
reactions (d) and (e), along with concomitant parasitic polymerization
reactions of the solvent at the Mg surface,\cite{Barile2014}  dictate
the process of electrolyte aging. 

 On the basis of reactions (d) and (e) of
Table~\ref{tb:equil_reactions}, we suggest that during the first few
electrochemical cycles of Mg deposition of a freshly prepared MACC
electrolyte, Al ions in solution (AlCl$_4^-$) are easily displaced,
thereby decreasing the initial coulombic efficiency of the electrolyte
as observed by Barile \emph{et al}\cite{Barile2014} ---this process is
called conditioning of the electrolyte (see discussion later). However,
during conditioning the presence of a chemical or electrical potential
promotes reaction (c) furhter towards the formation of
MgCl$^+$(AlCl$_4^-$) species, hence favoring Mg deposition over Al. The
concepts of electrolyte aging and conditioning will be clarified further
in the discussion section.

\subsection{$^{25}$Mg and $^{35}$Cl NMR properties of selected Mg$_x$Cl$_y$ structures}
To aid the interpretation of future NMR experiments on the MACC
electrolyte we computed the NMR isotropic shielding fingerprints of
$^{25}$Mg and $^{35}$Cl of selected MACC complexes. In general, changes
in charge density localization on different MACC complexes directly
alter the screening effects experienced by each NMR nucleus giving rise
to different NMR responses.

While Mg posses an NMR active nucleus, due to its low abundance
$^{25}$Mg requires expensive high field NMR instruments. Therefore, in
this analysis $^{25}$Mg data will be complemented by data on the more
abundant $^{35}$Cl nucleus. 

\begin{figure}[t]
\centering
\includegraphics[width=1.0\columnwidth]{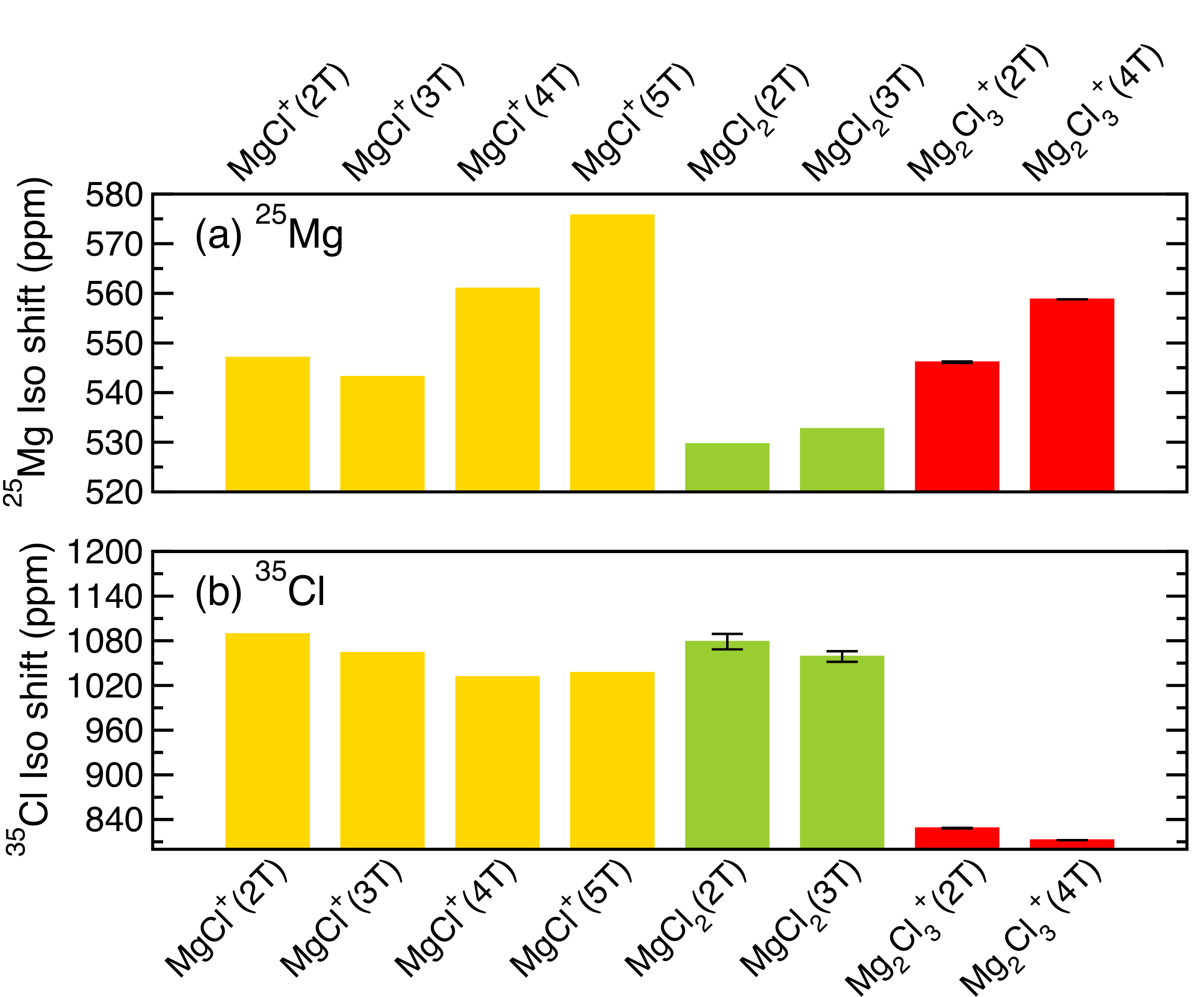}
\caption{\label{fig:mr_shifts} a) $^{25}$Mg and b) $^{35}$Cl NMR
isotropic shifts (in ppm) of relevant species magnesium-chloride complex
ions in the MACC electrolytes. Colors for different magnesium-chloride
complexes. Coordinating THF molecules indicated by T. Note that NMR data
are not shifted to $^{25}$Mg or $^{35}$Cl standard reference compounds.
Range of isotropic shifts for  compounds with more than one Mg or Cl
atoms is indicated by error bars.}
\end{figure}

Figure~\ref{fig:mr_shifts} shows the $^{25}$Mg and $^{35}$Cl NMR
isotropic shifts in THF for some relevant Mg$_x$Cl$_y$ clusters isolated
from the grand-potential phase diagrams of
Figs.~\ref{fig:mg_cl_thf_grandphasediag}
and~\ref{fig:mg_cl_thf_grand_drying}. 

The $^{35}$Mg and $^{35}$Cl isotropic shielding (of
Fig.~\ref{fig:mg_cl_thf_grandphasediag}) fall at very different absolute
values.  Note that NMR data of Fig.~\ref{fig:mr_shifts}a and
Fig.~\ref{fig:mr_shifts}b are not shifted to $^{25}$Mg and $^{35}$Cl
standard reference compounds. Our results indicate that $^{25}$Mg NMR
should be able to distinguish very well between charged Mg$_x$Cl$_y$
complexes (i.e.\ MgCl$^+$($n$T) and Mg$_2$Cl$_3^+$($n$T)) and neutral
species (MgCl$_2$($n$T)) in the MACC electrolyte. In addition we predict
that $^{35}$Cl NMR can discriminate between monomer and dimer species
(MgCl$^+$($n$T) and Mg$_2$Cl$^+_3$($n$T)), complementing $^{25}$Mg NMR
data.  Though the combination of $^{25}$Mg and $^{35}$Cl NMRs can
clearly differentiate between the stoichiometry of magnesium-chloride
complexes, our calculations suggest that it will be more difficult to
make conclusive claims on the effect played by the solvent (THF) with
NMR.  Furthermore, while the analysis of the Mg-Cl grand-potential PD
suggests that agglomeration of MgCl$^+\cdots$MgCl$_2$ is significantly
more likely to occur than the distinct dimer (Mg$_2$Cl$_3^+$, see
discussion later), the spectroscopic differences between the two species
may be subtle. For $^{25}$Mg data an increase of the isotropic shift is
observed for increasing THF coordination, see for example the trend for
MgCl$^+$($2$T) to MgCl$^+$($5$T). Less pronounced is the $^{35}$Cl NMR
shift decrease as a response to an increase of the THF coordination
number (see Fig.~\ref{fig:mr_shifts}b). 

\section{Discussion}
\label{sec:discussion}
In this work, the stable species present in the MACC electrolyte are
predicted using CMD and \emph{ab initio} calculations.

Although this investigation provides an important understanding of the
composition of the MACC electrolyte, it deals with bulk MACC solution
and does not explicitly account for: \emph{i}) the effect of the anode
and cathode surfaces, \emph{ii}) the existence of parasitic chemical
reactions that might alter the electrolyte composition, and \emph{iii})
the presence of impurities.

From the formation energies and grand-potential phase diagrams of
magnesium-chloride complexes we demonstrated that only two major Mg(Al)
species are present at equilibrium conditions in MACC, namely the
neutral MgCl$_2$(AlCl$_3$) and electro-active MgCl$^+$(AlCl$_4^-$). We
demonstrate that larger Mg$_x$Cl$_y$ units such as dimer and trimer are
not stable, though they might become accessible at room temperature by
changing the solvent conditions (drying/crystallization).
Polymerization of THF by AlCl$_3$ is also possible\cite{Barile2014,
Canepa2015}, and has the effect of decreasing the solvating capabilities
of THF towards the species in solution (MgCl$^+$ and MgCl$_2$).
Therefore THF-polymerization represents an alternative mechanism to
emulate drying conditions in solution and stabilize the dimer species.
In drying conditions achieved with crystallization procedures, the dimer
Mg$_2$Cl$_3^+$ has been successfully isolated,\cite{Pour2011} but
results in a electrochemically inert solution when redissolved in THF.
Benzmayza \emph{et al.},{\cite{Benzmayza2013}} speculated that the lack
of solvent in certain electrochemical conditions, for example when
MgCl$^+$ and MgCl$_2$ approach the anode surface, is responsible for the
formation of the dimer species. These experimental observations are
consistent with our theoretical findings suggesting that the operation
of the MACC electrolyte is ascribed to its simple chemical
structure/composition, and regulated by uncomplicated equilibria.

Interestingly, previous theoretical investigations of the monomer and
dimer coordination in THF,\cite{Wan2014} have demonstrated that the
symmetry of the dimer is largely perturbed by the THF solvent, forming
an open structure that resembles an isolated magnesium chloride molecule
interacting with a dangling monomer, i.e.\  MgCl$_2\cdots$MgCl$^+$.
Combining these observations, we speculate that the dimer Mg$_2$Cl$^+_3$
originates from the agglomeration of MgCl$_2$ available in solution and
MgCl$^+$. Under conditions of drying/crystallization, similar
agglomeration mechanisms can explain the formation of larger order
magnesium-chloride structures (e.g.\ trimer and polymeric units), which
have been speculated to exist.\cite{Barile2014} To this end, we have
computed useful $^{25}$Mg and $^{35}$Cl NMR fingerprints of the stable
and unstable MACC species. 

Our findings also shed light on the coordination of inorganic aluminum
magnesium-chloride complexes. In line with preliminary experimental and
theoretical work,\cite{Nakayama2008,Wan2014,Liu2015a} we demonstrate
that magnesium-chloride salts in THF solutions cannot fulfill the
typical 6-fold coordination of Mg$^{2+}$ in solids, but always prefer
lower coordination numbers (e.g.\ 4-fold for the monomer
MgCl$^+$(3THF)).  According to the vast organic
literature,{\cite{Guggenberger1968,Garsta2004,Pirinen2013}}  Grignard
reagents' MgXR$_2$ (with X = Cl, Br) and halides salts (MgCl$_2$ and
MgBr$_2$) in THF are typically found 4-fold coordinated, and confirm our
findings. Compared to multi-dentate linear glymes (e.g. diglyme and
tetraglyme) the ability for THF to coordinate ions is limited by the
bulkier structure of the ring, and this has been also demonstrated
experimentally and computationally by Seo \emph{et  al.}{\cite{Seo2014}}
Moreover, the coordination environment in the crystalline state does not
necessarily reflect the coordination in the liquid
phase.{\cite{Ansell1997,Wernet2004}}

 A closer analysis of our data shows that the stable Mg coordination
number increases as a function of the Mg-Cl complex size from monomer to
trimer. In a recent study, some of us attested that lower Mg$^{2+}$
coordination numbers decreases the desolvation energy required to shed
the solvent during plating and stripping.\cite{Canepa2015} We speculate
that the larger Mg$^{2+}$ desolvation energy for bigger Mg-Cl complexes
(e.g.\ dimer and trimer) can inhibit the delivery of fresh Mg$^{2+}$ at
the Mg-anode during plating. 

By identifying the principal species of the MACC electrolyte at
equilibrium, MgCl$^+$(3THF), MgCl$_2$(2THF), AlCl$_4^-$, and
AlCl$_3$(THF), we can explain the phenomenological effects observed in
the MACC electrolyte under electrochemical cycling.  A thermodynamic
analysis of the bulk electrolyte properties suggests that the
equilibrium between MgCl$^+$ and MgCl$_2$ (and AlCl$_4^-$ and AlCl$_3$)
in THF tends towards a solution dominated by charged MgCl$^+$ (and
AlCl$_4^-$) species, (see reaction (c)
Table{~\ref{tb:equil_reactions}} corrected by the Debye-H\"{u}ckel
model), which provides the appropriate conditions for ion conductivity.
The $\Delta E$s calculated for each equilibria dictate the activity
ratio between MgCl$^+$ and MgCl$_2$ that impacts the number of charge
carriers (MgCl$^+$) available in solution, and ultimately impacts the
ionic conductivity of the MACC electrolyte.  In MACC AlCl$_4^-$
functions as a shuttle replenishing Cl$^-$ ions (at the anode surface)
during Mg stripping (at the anode);\cite{Canepa2015}  reaction (c) of Table~\ref{tb:equil_reactions} suggest that the ratio between
AlCl$_4^-$ and AlCl$_3$ is large, hence  allowing the complex
dynamics of Mg stripping and dissolutions.

The availability of MgCl$^+$ in solution is not only controlled by
reaction (c) but also depends on the low solubility of MgCl$_2$ in THF
(see reaction (a)). Liao \emph{et al.}{\cite{Liao2015}} demonstrated
that the solubility of MgCl$_2$ can increase dramatically provided the
presence of Cl$^-$ acceptors in solution. While AlCl$_3$ seems
appropriate (as demonstrated  by reaction (c)),  other Cl$^-$ ions
acceptors can be introduced as ``additives`` (e.g. Mg(HMDS)$_2$)
promoting large quantities of MgCl$^+$ in solution. 

However, by using the Debye-H\"{u}ckel corrected $\Delta$E of
reaction (c)  we find that a significant concentration of charge
carriers is still available in solution. For example using  a typical
concentration of 0.5~M for AlCl$_3$, and assuming that the maximum
activity of soluble MgCl$_2$ in THF is~7.8$\times$10$^{-4}$~M (set by
reaction (a)),   we expect a concentration of MgCl$^+$(3THF) in solution
to be $\sim$~0.092~M.  Notably, for this  concentration we could derive,
using the Kohlrausch's  law for weak electrolytes,  an ionic
conductivity of $\sim$~1.96~mS~cm$^{-1}$, which is in very good
agreement with the experimental value measured by Doe \emph{et al.}
($\sim$~2~mS~cm$^{-1}$) for a fully conditioned
electrolyte.{\cite{Doe2014}} See Supplementary Information for the full
derivation of the ionic conductivity. 

The Al-Cl-Mg-THF phase diagram does not indicate the formation of stable
${\rm AlCl}_4^-\cdots {\rm Mg}_x{\rm Cl}_y$ ionic couples, though some
of these clusters might be accessible within small energy windows
(0.064~$-$~0.088~eV) with further repercussions on electrolyte
conductivity. In general, the small dielectric constant of THF
($\sim$~7.58) and glymes favor the formation of ionic couples, an
indication that the next generation of solvents for Mg-ion batteries
requires solvent with better screening properties. 

\begin{figure*}[!ht]
\centering
\includegraphics[scale=0.85]{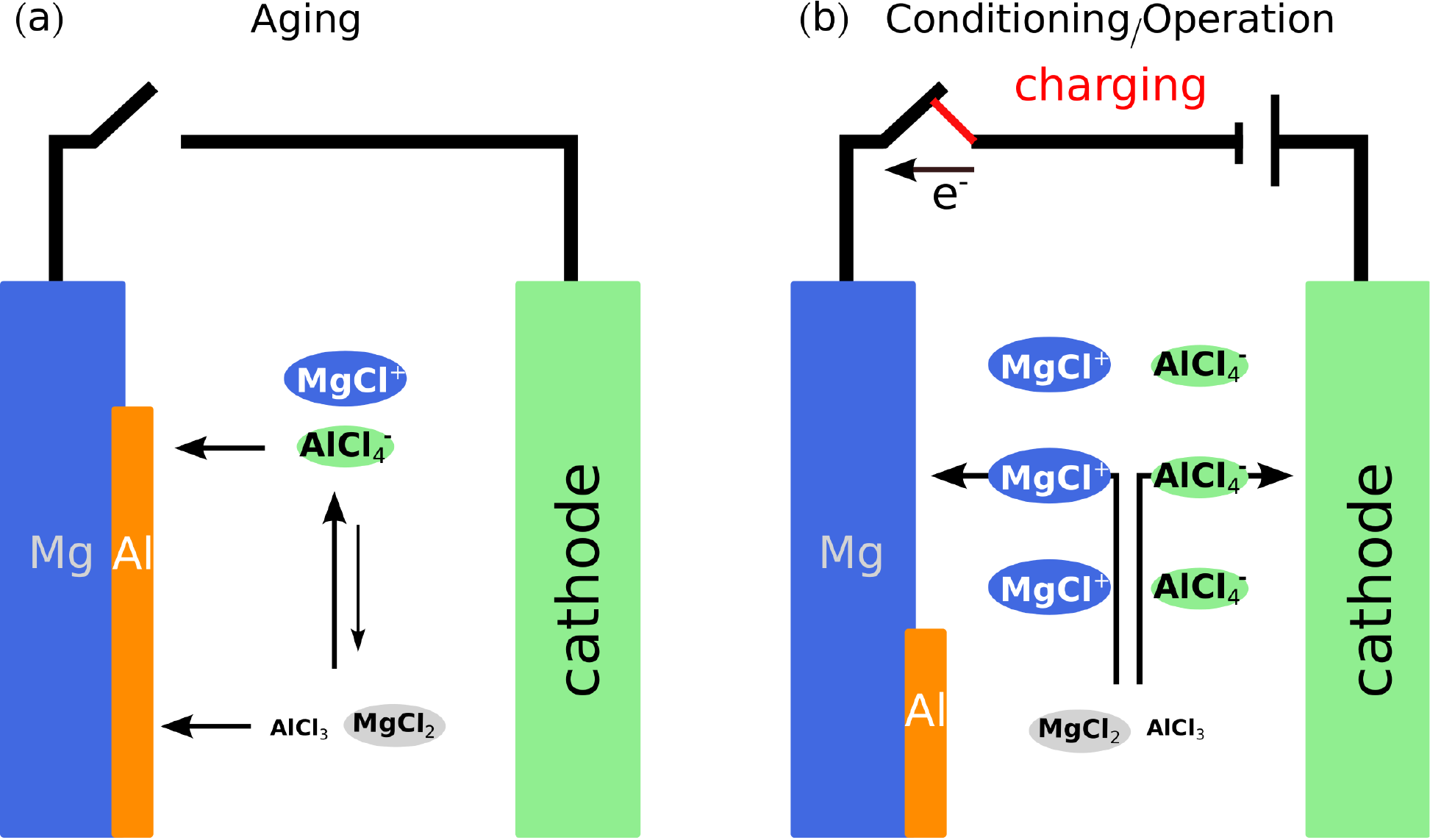}
\caption{\label{fig:agingcond_schematic} Schematic of the processes of
aging (a) and conditioning (b) of the MACC electrolyte in a battery
setup. Panel (a)  Al deposition at the Mg-anode electrode is depicted by
orange stripes and supported by reaction (d) and (e) of
Table~{\ref{tb:equil_reactions}}. Panel (b) shows the activity gradients
of MgCl$^+$ and AlCl$_4^-$ arising from the Mg plating process as well
as continuous regeneration of these species in the bulk solution.
During aging the species (i.e.\ MgCl$^+$, MgCl$_2$, AlCl$_4^-$, and
AlCl$_3$) in solution are distributed homogeneously in the electrolyte.
}
\end{figure*}

Although we do not explicitly consider the Mg-electrode, from the
reaction energy discussed in Table~\ref{tb:equil_reactions} we  provide
important considerations on the process of aging of the electrolyte.
Figure~\ref{fig:agingcond_schematic} summarizes the processes of aging
(a) and conditioning (b) of the MACC electrolyte.  From the equilibrium
between Mg-Cl-Al species in solutions and Mg/Al bulk metals, we
demonstrate that AlCl$_4^-$ ions in solutions are easily displaced
during the initial stages of Mg deposition. In fact, Al deposition at
the anode is ensured by a small Al reduction potential ($\sim$~--1.67~V
vs.\ NHE)  compared to Mg ($\sim$~--2.35~V vs.\ NHE) setting a
thermodynamic driving force for the process of aging. Finally, further
aluminum depletion from the solution upon electrolyte resting which
impacts the amount of MgCl$^+$ in solution, could be one of the causes
of aging of the MACC electrolyte.  Additionally, parasitic
polymerization reactions of the solvent have also been
speculated\cite{Barile2014} to be the source of electrolyte aging. 

Under open circuit conditions (battery at rest), the reaction at the
anode/electrolyte interface is largely controlled by the activity of
MgCl$^+$ species  available since reaction (c)  dominates the
composition of the electrolyte (see Table~{\ref{tb:equil_reactions}}).
Reactions (d) and (e) favor the formation of Al deposition under open
circuit, leading to AlCl$_4^-$  (and MgCl$^+$) depletion form the
solution, resulting in the aging of the electrolyte.  However, when an
aged electrolyte is subjected to charging, the presence of an applied
potential drives Mg deposition on the anode, resulting in not only
setting a concentration gradient of MgCl$^+$ (AlCl$_4^-$) from the bulk
towards the anode (cathode) but also a continuous regeneration of
MgCl$^+$ in the solution. After aging, the electrolyte will require a
few charge-discharge cycles before the composition in the solution is
stabilized and the charged species (MgCl$^+$ and AlCl$_4^-$) are
abundantly present leading to smooth Mg deposition/stripping.
Therefore, the state of conditioning in the electrolyte represents a
transition between the Al-deposition regime (aging) and the Mg
deposition/stripping regime during regular battery operation. Barile
\emph{et al.}\cite{Barile2014} have estimated that about 100
electrochemical cycles are needed to condition the MACC electrolyte.
This explains why low coulombic efficiencies of fresh MACC solutions
have been attributed to Al deposition during the initial electrochemical
cycles. SEM-EDS measurements of a Pt electrode that underwent Mg
deposition during electrolyte conditioning showed large quantities of
permanently deposited Al,\cite{Barile2014}  corroborating our modeling
results. 

Barile~\emph{et al.}{\cite{Barile2014}} suggested that the MACC
electrolyte is conditioned when the Mg/Al molar ratio in solution is
$\sim$~2.6:1, from which they concluded that dimer species must be
present in the electrolyte. However, our grand-potential phase diagrams
(see Fig.~{\ref{fig:mg_cl_thf_grandphasediag}} and
Fig.~{\ref{fig:mg_cl_thf_grand_drying}}) indicate that dimer species are
unlikely to be present in the electrolyte at equilibrium, but only
become accessible when drying or crystallizing the electrolyte. The
Mg/Al ratio observed experimentally ($\sim$~2.6:1) for conditioned
electrolytes can alternatively stem from the presence of agglomerates
MgCl$^+\cdots$MgCl$_2$ (the only stable species in solution) instead of
distinct dimer ions ---MgCl$^+\cdots$MgCl$_2$ clusters have been
isolated previously using \emph{ab initio} MD on a similar
electrolyte.{\cite{Wan2014}}

\section{Conclusions}
\label{sec:conclusions}
With the intention of elucidating the structural composition of the MACC
electrolyte, we carried out \emph{ab initio} calculations and
classical molecular dynamics simulations on more than a hundred
molecules and ions that could be structurally and functionally relevant
for this electrolyte.  We find that only MgCl$^+$, MgCl$_2$, AlCl$_4^-$
and AlCl$_3$ are stable constituents of  the electrolyte. The
thermodynamic analysis of the MACC composition excludes the presence of
multimeric Mg$_x$Cl$_y^+$ units such as dimer and trimer under
equilibrium conditions.  These species can be stabilized under
conditions of solvent drying. 

Equilibrium between the MACC species (i.e\ MgCl$^+$ MgCl$_2$, AlCl$_4^-$
and AlCl$_3$) in liquid THF and Mg and Al metals suggests that Al is
easily displaced from the solution during early Mg deposition cycles.
This effect reduces the electrolyte coulobmic efficiency providing an
explanation for the process of aging. In general, Al deposition on
Mg-metal is always favored and leads to the more complex issue of
electrolyte aging. We explain conditioning as the process which promotes
the stabilization of charged species (MgCl$^+$ and AlCl$_4^-$) in
solution due to a potential (chemical or applied), enabling Mg smooth
deposition/stripping.

Computation of the NMR shifts of the relevant MACC species shows
distinct $^{25}$Mg and $^{35}$Cl NMR signatures for monomer, dimer and
MgCl$_2$, concluding that \emph{in-situ} NMR can clarify the composition
of the MACC electrolyte as well as transformation of the MACC solution
occurring during aging and conditioning of the electrolyte. Our
analysis indicates that MgCl$_2$ is sparingly soluble in THF, but its
solubility can be increased by introducing Cl$^-$ acceptors.
Finally, the computational strategy adopted in this investigation is
readily applicable in a high-throughput fashion to  study other liquid
media, specifically to progress the understanding of liquid
electrolytes, and to screen for new electrolytes for  the next
generation of rechargeable batteries. 

\begin{acknowledgements}
This work was fully supported as part of the Joint Center for Energy
Storage Research (JCESR), an Energy Innovation Hub funded by the U.S.
Department of Energy, Office of Science, and Basic Energy Sciences. This
study was supported by Subcontract 3F-31144. We also thank the National
Energy Research Scientific Computing Center (NERSC), a DOE Office of
Science User Facility supported by the Office of Science of the U.S.
Department of Energy under Contract No. DE-AC02-05CH11231, for providing
computing resources.  PC and SJ are thankful to Dr.\ Xiaohui Qu at the Lawrence
Berkeley National Laboratory for numerous suggestions.  All authors are
also indebted to Dr.\ Kevin R. Zavadil at Sandia National Lab, Dr.\
Christopher Barile and Prof.\ Andrew Gewirth at University of Illinois
Urbana-Champaign for stimulating discussions.
\end{acknowledgements}

\bibliography{biblio}

\end{document}